\documentclass[aps,twocolumn]{revtex4-2}
\usepackage{siunitx}
\usepackage{booktabs}
\usepackage{amsmath}
\usepackage{graphicx} 
\usepackage{gensymb} 
\usepackage{soul}
\usepackage{aasmacros}

\usepackage[final]{changes}

\begin{document}
\title{Two-dimensional Synchrotron Beam Characterisation from a Single Interferogram } \date{17 May 2024}

\author{Bojan Nikolic}
\email[E-mail:]{bn204@cam.ac.uk}
\affiliation{Cavendish Laboratory, University of Cambridge, Cambridge CB3 0HE, UK}

\author{Christopher L. Carilli}
\affiliation{National Radio Astronomy Observatory, P. O. Box 0, Socorro, NM 87801, US}

\author{Nithyanandan Thyagarajan}
\affiliation{Commonwealth Scientific and Industrial Research Organisation (CSIRO), Space \& Astronomy, P. O. Box 1130, Bentley, WA 6102, Australia}

\author{Laura Torino}
\email[E-mail: ]{ltorino@cells.es}
\affiliation{ALBA - CELLS Synchrotron Radiation Facility\\Carrer de la Llum 2-26, 08290 Cerdanyola del Vallès (Barcelona), Spain}

\author{Ubaldo Iriso}
\affiliation{ALBA - CELLS Synchrotron Radiation Facility\\Carrer de la Llum 2-26, 08290 Cerdanyola del Vallès (Barcelona), Spain}

\begin{abstract}
  Double-aperture Young interferometry is widely used in accelerators
  to provide a one-dimensional beam measurement. We improve this
  technique by combining and further developing techniques of
  non-redundant, two-dimensional, aperture masking and
  self-calibration from astronomy. Using visible synchrotron
  radiation, tests at the ALBA synchrotron show that this method
  provides an accurate two-dimensional beam transverse
  characterisation, even from a single 1~ms interferogram. The
  {non-redundancy of the aperture mask in the} technique
  {enables it to be} resistant to {spatial} phase fluctuations that
  might be introduced by vibration of optical components, or in the
  laboratory atmosphere.
\end{abstract}

\maketitle

\section{Introduction}

Two-dimensional (2D) beam size measurements in the transverse plane
are of fundamental importance to quantify the performance of
accelerators. They allow for the characterization of the emittance of the
particle beam, a key parameter defining an accelerator. The
transverse distribution of particles is well characterized by a 2D
Gaussian~\cite{Sands:1969lzn} and parameterised by major axis, minor axis, and the tilt angle.

In synchrotron light sources, noninvasive beam size measurements
exploiting the Synchrotron Radiation (SR) emitted by the electron beam
are mainly performed via direct imaging or interferometry
techniques~\cite{Kube:Dipac07}. Among the direct imaging techniques,
the X-ray pinhole~\cite{Elleaume:km0006} is widely used in synchrotron
light sources as it directly provides a 2D image of the source.  {A
recent design study for X-ray pinhole cameras can be found in
\cite{PhysRevAccelBeams.27.032802}.}

At longer wavelengths the direct imaging is limited by diffraction
and interferometric techniques are preferred. Among them, the
two-aperture interferometer \cite{Mitsuhashi:1998em} is widely used in
accelerators, but it only provides a one dimensional measurement in
the direction of the orientation of the apertures.  Full 2D beam
reconstruction can be obtained by a series of measurements, rotating
the aperture mask between
measurements~\cite{PhysRevAccelBeams.19.122801}, but at least four
orientations are needed. A four-aperture square
interferometric mask layout \cite{Masaki:wr2006} has been used to obtain a 2D source size from a single interferometric measurement, but such a mask layout suffers from decoherence due redundant sampling in Fourier
space~\cite{Carilli24}. Moreover, the non-uniformity of aperture
illumination across the mask due to the intrinsic distribution of the
synchrotron radiation can irrecoverably corrupt the
results~\cite{Novokshonov:2017qsa}. {Redundant two-dimensional  mask interferometry at X-ray wavelengths has also been proposed \citep{Li:23} but similarly it will not be able to recover the relative phases and illuminations of the individual apertures.}

{In the last years several techniques are arising to guarantee high resolution and fast measurement needed from the upcoming 4th generation of synchrotron light sources \citep{PhysRevAccelBeams.25.052801, PhysRevLett.131.185001, Curcio2024}.}

{In this paper we propose the combination of Non Redundant Aperture (NRA) mask interferometry in combination to gain fitting \cite{1987Natur.328..694H, 1980SPIE..231...18S}, two technique broadly used in astronomy separately and united here for the first time. This combination provides a full transverse characterization of the electron beam in a single acquisition, using the visible part of the synchrotron radiation. The main components of the optical setup are out of vacuum and data is acquired using a standard CCD. The analysis can be performed online and results can be used in a feedback loop to guarantee a stable emittance in both planes}

In this paper, we present the technique in detail and use it to characterize the ALBA electron beam.

{One alternative method of two-dimensional constraint on the
  synchrotron beam size at X-ray wavelengths is by measuring the power
  spectrum of speckle formed by a random scattering medium
  \citep{PhysRevAccelBeams.25.052801}, an approach also used in
  astronomy \citep{1970A&A.....6...85L}. Like our approach, speckle
  also depends on the interference of the synchrotron radiation but
  unlike the present method, speckle power spectrum only contains
  information about the amplitude of the coherence function and
  requires operation with X-rays in a vacuum for sufficient
  resolution. }

{While angular resolution of direct imaging at visible
  wavelengths is limited, it is possible to improve the achieved
  transverse resolution by placing a high-accuracy lens physically
  close to the radiation source \cite{PhysRevLett.131.185001}. This
  approach, however, needs to deal with the intense X-ray flux, be
  implemented largely in vacuum, and is difficult in the next-generation synchrotrons which have narrow radiation extraction
  angles due to space constrains.}

{Finally, in some cases it is possible to use detailed
  measurement of the angular distribution of the spectrum of the
  emitted synchrotron radiation to infer the dimensions of the
  particle beam\cite{Curcio2024}. }

\section{Experimental setup}
\label{sec:experimental-setup}

The experiments were carried out at the ALBA synchrotron facility.  The laboratory optical bench setup and the imaging camera were similar to the
two-aperture interferometry described by
\cite{PhysRevAccelBeams.19.122801}, except that we use new
multi-aperture masks and different integration times.

\begin{figure}
\centering 
\centerline{\includegraphics[scale=0.41,trim=1cm 14cm  0 5cm]{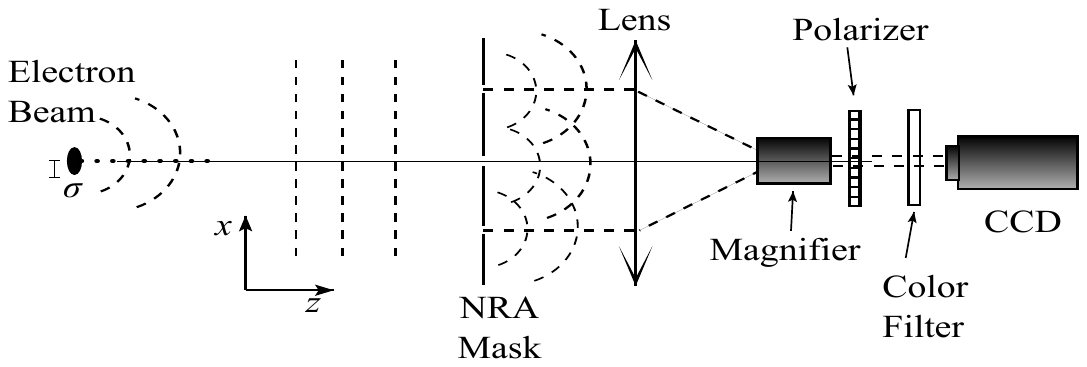}}
\caption{{Schematic diagram of the experimental setup.}}
\label{fig:expset}
\end{figure}

{ The visible part of the synchrotron radiation, emitted from a
  bending magnet, is extracted by an in-vacuum mirror positioned at a
  distance of \SI{8.64}{m}. The mirror captures only the upper lobe of the
  radiation, cutting the wavefront \SI{10}{mm} above the orbit
  plane. The light then passes through a vacuum window and is transported
  to the beamline optical table via 7 additional mirrors, covering a
  total optical path from the source of approximately \SI{15}{m}
  \cite{AAD-FE-DI-VMIR-01}. After diffraction by the NRA mask and
  focusing with a \SI{0.5}{m} focal length lens, the image is magnified,
  filtered using a $538\pm$\SI{10}{nm} color filter, polarized to
  select $\sigma$ polarization, and finally captured by a CCD camera.
  A sketch of the experiment is illustrated in Fig.~\ref{fig:expset}.}

We present source size measurements made using a five-aperture mask
designed such that the vector (`baseline') between every pair of
apertures is unique (a `non-redundant' mask). 
This means that each spatial frequency in the Fourier domain is measured by only a single pair of apertures in the plane of the mask, thereby avoiding decoherence that would occur for redundant aperture measurements in the presence of phase errors. This mask was an adaptation of the non-redundant array in \cite{Gonzalez:11}, with the five apertures selected to maximize the longer baselines (given the source is only marginally resolved) while fitting within the illuminated area.

\begin{figure}
\centering 
\centerline{\includegraphics[scale=0.7,trim=100 220 400 100,clip]{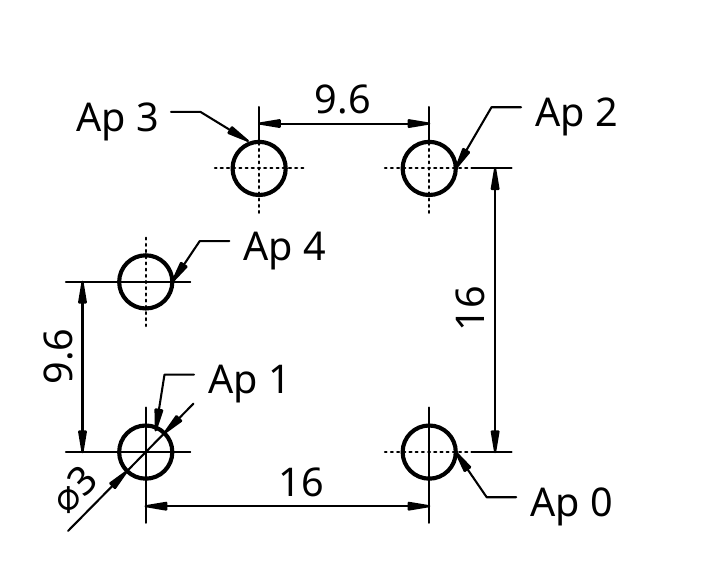}}
\caption{Drawing of the non-redundant mask with aperture labels as
  used for labelling baselines. (All dimensions in millimetres).}
\label{fig:mask3mm}
\end{figure}

The geometry of the mask is shown in Fig.~\ref{fig:mask3mm}.  The
hole sizes were 3mm in diameter and the mask was machined out of
aluminium to high accuracy in the ALBA mechanical workshop (better than 0.1\,mm).

\section{Measuring visibilities}

Data are acquired as CCD two-dimensional arrays of size
$1296\times966$. We first remove the constant offset which is due to a
combination of the bias and the dark current. We use a fixed estimate
of this offset obtained by examination of the darkest areas of the CCD, as well as
from the FFT of the image. Errors in this procedure accumulate in the
central part of the FFT and contribute to the overall uncertainty of
the beam reconstruction.

Next we pad and center the data so that the centre of the Airy
disk-like envelope of the fringes set by the hole diameter is in the
centre of a larger two-dimensional array of size $2048 \times
2048$. To find the correct pixel to center to, we first filter the
acquired image with a wide (50 pixel) Gaussian kernel, then select the
pixel with highest signal value. The Gaussian filtering removes the
fringes leaving the remaining signal approximately corresponding to
the fringe envelope. Without the filtering, the particular central
pixel selected would be affected by the fringe position and the photon
noise, rather than the envelope.

\begin{figure}
\centering 
\centerline{\includegraphics[scale=0.5]{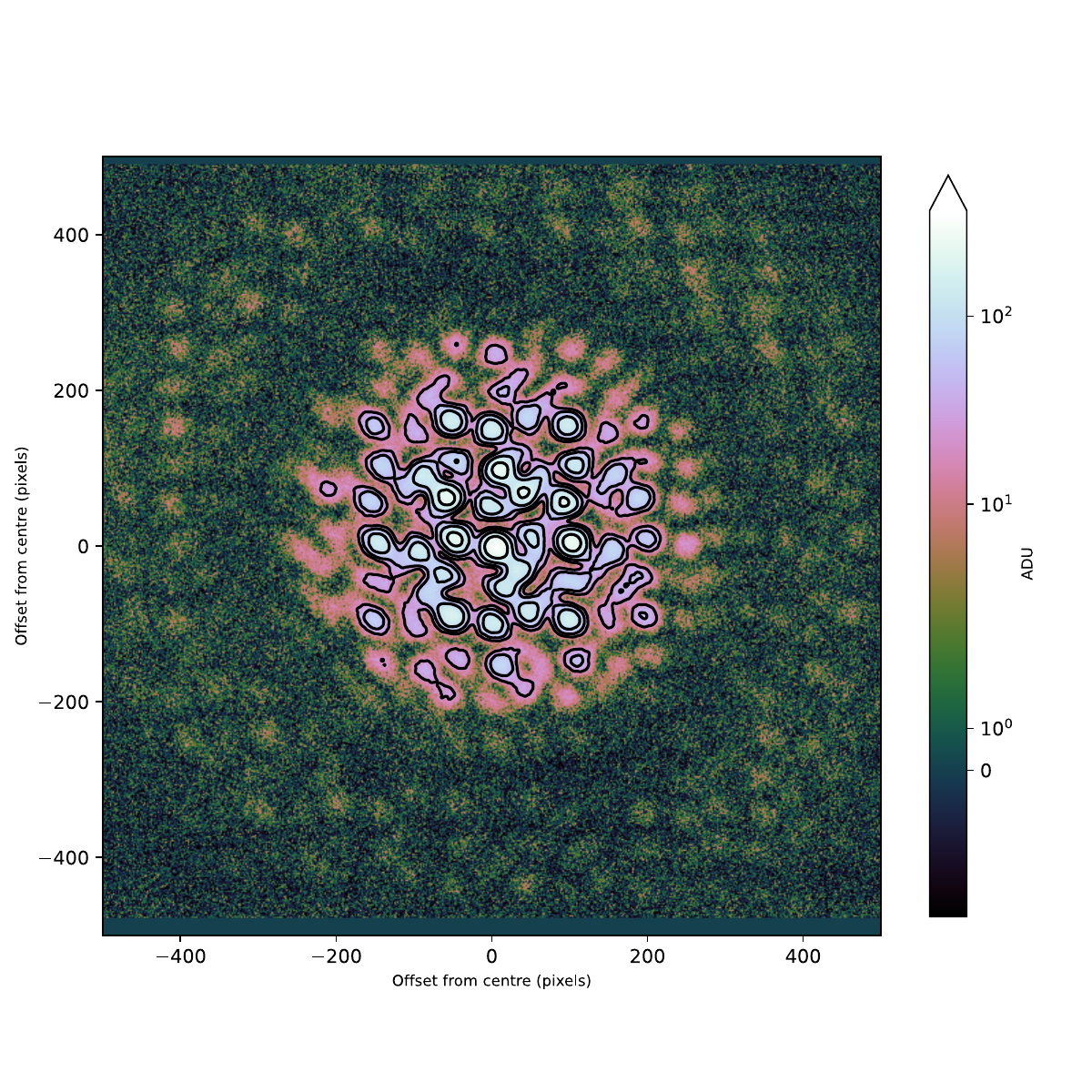}}
\caption{Interferogram with five 3\,mm-diameter holes and 1\,ms
  integration time. The CCD image is bias-subtracted and centered. Contours are power-law in $2^{-n}$.
  }
\label{fig:3mminterf}
\end{figure}  

An example CCD frame processed in this way is shown in
Fig.~\ref{fig:3mminterf}. The primary and first side lobe of the Airy
disk envelope of the fringes can be seen. The measurement has high
signal-to-noise with the fringes even in the first side-lobe clearly
discernible above the noise.

\begin{figure}
\centering 
\centerline{\includegraphics[scale=0.45]{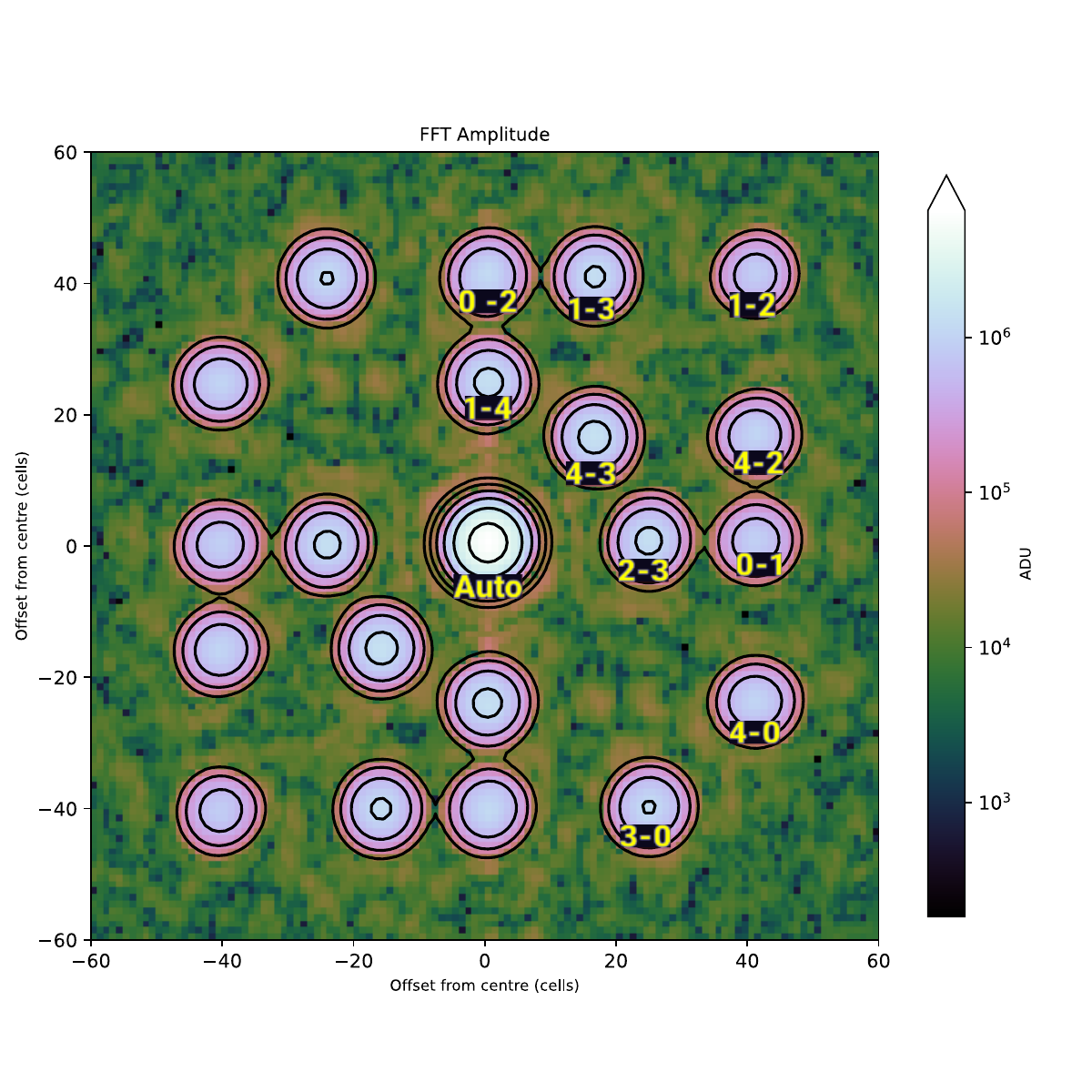}}
\caption{Amplitude of the FFT of the sample interferogram in
  Figure~\ref{fig:3mminterf}, contours drawn at $2^{-n}$ intervals
  relative to image maximum. Labels identify peaks of amplitude with
  the pair of apertures forming the baseline of visibility that the
  peak represents.}
\label{fig:3mmfft}
\end{figure}

To calculate the visibilities, {we use the Fraunhofer
  diffraction integral, and the result that the Fourier transform of
  an intensity image is the auto-correlation of the field in the
  aperture plane}.  We therefore compute the two-dimensional Fourier
transform of the padded CCD frame using the FFT algorithm. Amplitude
of an example Fourier transform of an ALBA image is shown in
Figure~\ref{fig:3mmfft}. Distinct peaks can be seen in the FFT
amplitude corresponding to each baseline (i.e., vector) between the
apertures in the mask.

We extract the {visibilities} on each of the baselines by
calculating the complex sum of pixels within a circular aperture of
radius 7\,pixels, centered at the calculated position of the
baseline. With the array padding used here, 1\,mm on the mask
corresponds to 2.54 pixels in the Fourier transformed interferogram
and the correlated signal is expected up to twice the hole radius so
the 7 pixel aperture captures almost all of the signal while
minimising noise and systematics.  The resulting complex sum value
contains both the amplitude and phase of the visibility.

\section{Fitting the coherences}

The measured visibilities $ \|V_{ij}\|$ are a function of both the
degree of coherence {$\gamma(u_{ij}, v_{ij})$} and the
amplitudes of the electromagnetic radiation passing through each of
the apertures {$\|G_i\|,\, \|G_j\|$, where $i$ and $j$ are 
  indices of the apertures:} 
\begin{equation}
  \label{eq:visdef}
    \|V_{ij}\|  =  \gamma(u_{ij}, v_{ij}) \|G_i\| \|G_j\|.
\end{equation}
Since it is not possible to ensure the intensity of radiation through
the apertures are all the same, our strategy is to simultaneously fit
for both the coherences as well as the`'gains' that correspond to the
square root of intensity of the incident radiation at each of the
apertures. If this fitting is not performed, errors introduced by
non-uniform illumination are greater in the multi-aperture
interferometry than in the two-aperture case because we only have a
constraint on the total power of the radiation through all the
apertures, not on the particular two apertures forming each
visibility. This means that without fitting, a fractional difference
in illumination of $\epsilon$ leads to an error in coherence of order
$\epsilon^2$ in the two-aperture case but order $\epsilon$ in the
multi-aperture case.  {Furthermore, while with two holes it
  possible to position them so that difference in illumination is
  minimised, this is impractical with five holes}.

{ The central peak of the Fourier transform brings information
  on the auto-correlation and adds another constrain to the fit being
  the total signal being the sum of all the gains:
  \begin{equation}
    \label{eq:totpower}
    \|V_{\rm auto}\|  =  \sum_i  \|G_i\|^2.
\end{equation}
}

\subsection{Parametrisation of the model for the coherence}

By {suitable approximations of} the van Cittert–Zernike theorem (cite TMS 2017), the coherence can be very well
represented by the Fourier transform of the source intensity
distribution. Working on the assumption that the source is a
multi-variate Gaussian~\cite{Sands:1969lzn}, then the coherence is
also a multi-variate Gaussian, which is what we use as the model. We
fit a parametrisation in terms of the overall width ($\sigma$)
and the distortion in the vertical-horizontal directions
($\eta$) and diagonal ($\rho$) directions:
\begin{align}
  \gamma(u,v) =&\exp[- \frac{(u^2 + v^2) + 2 \rho (u v) + \eta(u^2 - v^2)}{2 \sigma^2} ]
\end{align}
{and the independent variables $u$ and $v$ are coordinates of a
  displacement in the plane of the aperture mask. In this
parametrisation in the plane of the aperture mask the major axis
$\sigma_u^0$, minor axis $\sigma_v^0$, and the tilt angle $\phi$ of the
major axis $\phi$ as measured from $u$ direction are:
\begin{align}
  \label{eq:majminphi}
   \sigma_u^0=&\sigma \left(1-\sqrt{\eta^2 + \rho^2}\right)^{-\frac{1}{2}} \\
   \sigma_v^0=&\sigma \left(1+\sqrt{\eta^2 + \rho^2}\right)^{-\frac{1}{2}} \\
  \phi =& \tan^{-1}\left(\frac{\rho}{\eta - \sqrt{\eta^2+\rho^2}}\right).
\end{align}}

This is a convenient parametrisation for fitting as it avoids
trigonometric functions and has low covariance of parameters in
typical situations.

\subsection{Likelihood function}

Under the assumption of normally distributed uncorrelated errors in
the visibility measurements the likelihood is:
\begin{align}
  - \log(L) = & \frac{ \|V_{\rm auto}^{\mathrm{obs}} - V_{\rm auto}^{\mathrm{mod}} \|^2}{\xi_{\rm auto}^2} + \sum_{i=0}^{N-2} \sum_{j=i+1}^{N-1} \frac{ \|V_{ij}^{\mathrm{obs}} - V_{ij}^{\mathrm{mod}} \|^2}{\xi_{\rm vis}^2} 
\end{align}
where, $N$ is the number of apertures in the mask (5 in this case),
$\|V_{ij}\|$ are the visibilities (Eq~\ref{eq:visdef}),
$\|V_{\rm auto}\|$ is the total power (Eq~(\ref{eq:totpower}),
{and $\xi_{\rm auto}$ and $\xi_{\rm vis}$ are the estimates of
  uncertainty for the total power (or autocorrelation) and visibility
  measurements, respectively. We estimated the uncertainty,
  $\xi_{\rm vis}$, by making multiple samples in the Fourier
  transformed plane (i.e., the data shown in Figure~\ref{fig:3mmfft})
  at positions which were not close to the visibility peaks. These
  parts of the visibility plane for which the input light is blocked
  by the optical system (the objective or the aperture mask) and hence
  contain only the random uncorrelated noise produced within the
  CCD. Variance between these samples is a good estimate of the random
  uncertainty on the visibility measurement.  For $\xi_{\rm auto}$, we
  add an additional 1\% relative uncertainty due to systematic effects
  such as the bias subtraction that tend to accumulate in the total
  power measurement.}

\subsection{Minimisation procedure}

The best-fitting model was found by maximising the likelihood using
the Levenberg–Marquardt algorithm with a numerically evaluated
Jacobian. The fitting processing time is dominated by the Fourier
transform computations and can be completed in around 25ms on a
server-class CPU (dual Intel Xeon 8276), faster than the typical CCD
read-out time. This means that, although in present experiments we
analysed the data offline, the technique could be used for a system
implemented for beam characterisation with low and predictable
latencies typically dominated by the fixed latency of the CCD readout,
i.e., near real-time measurement.

\section{Results}

We obtained a number of measurement runs on 2023-06-14 with 3\,mm and
5\,mm apertures, and with 1\,ms and 3\,ms integration times.  Each run
consisted of 30 CCD frames separated by one second. Data with
3\,mm apertures and 1\,ms integration time have the least time
variation and highest apparent coherences indicating that it is of
best quality~\cite{Carilli24}. 

\subsection{Beam size measurement}
\label{sec:beam-size-meas}

Using the 1~ms integration time data, we analysed each frame to derive
both the illumination through each mask hole and the best fitting
values for the coherences. The coherences are then used to calculate the size of the major axis of the electron beam 
{ 
using:
\begin{align}
  \sigma_\mathrm{major} =  \frac{\lambda L}{2 \pi \sigma_v^0},
\end{align}
\noindent
where $\lambda=\SI{538}{nm}$ and $L=\SI{15}{m}$ as described in
Section~\ref{sec:experimental-setup}. An analogous equation holds for
the minor axis. The angle of the major axis of the beam is rotated by
90 degrees compared to the angle of the major axis of the coherence
function given in Eq~(\ref{eq:majminphi})}.

\begin{figure}
\centering 
\centerline{\includegraphics[scale=0.6]{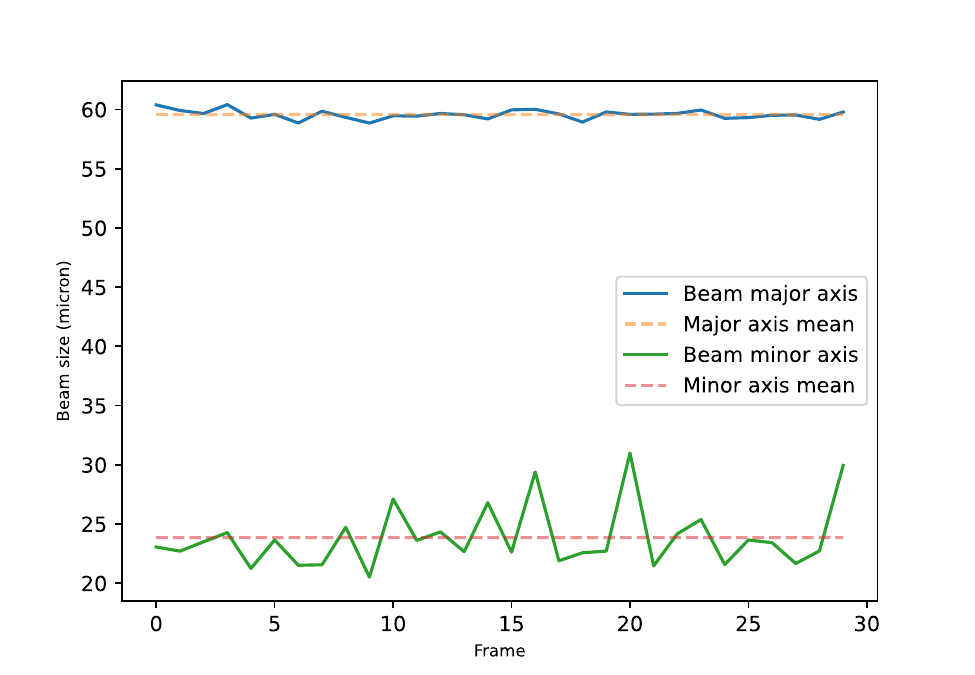}}
\caption{Time series of measured beam major and minor axes, $\sigma$, from 30 interferogram frames each separated by one second. Dash lines are mean values.}
\label{fig:3mmbeamts}
\end{figure}

\begin{figure}
\centering 
\centerline{\includegraphics[scale=0.6]{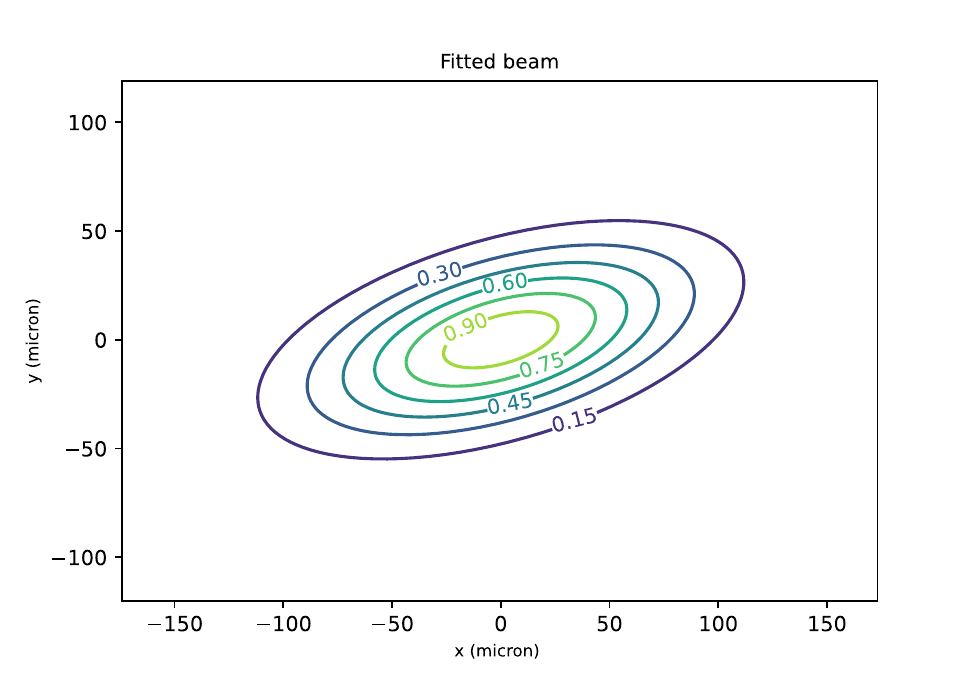}}
\caption{ Full 2D beam reconstruction from the average of 30 frames each 1\,ms integration time.}
\label{fig:3mmbeamrecon}
\end{figure}

The time series of these measurements,
shown in Fig.~\ref{fig:3mmbeamts}, show high stability and repeatability, with an rms scatter of
\SI{0.4}{\micro\metre} and \SI{2.6}{\micro\metre} in beam major and minor axes, respectively, and 0.9\degree\ in the angle.
The final beam derived is the average of the time series, which is shown in
Fig.~\ref{fig:3mmbeamrecon}.  The mean estimate of the axes sizes is
$\sigma_\mathrm{major} = \SI{59.6}{\micro\metre} \pm
\SI{0.1}{\micro\metre} $ and
$\sigma_\mathrm{minor} = \SI{23.8}{\micro\metre} \pm
\SI{0.5}{\micro\metre}$. The error on the minor axis is larger
because of its significantly smaller size, and also because the
available baselines are shorter, being limited by the size of the
illuminated region.  The angle the major axis makes with the
horizontal is $15.9\degree \pm 0.2\degree$.

\subsection{Comparison with other methods}

On the same day we also obtained measurements using rotated
two-aperture SRI as described by
\cite{PhysRevAccelBeams.19.122801}. This method has previously been shown
to be reliable and uses a different fitting procedure to obtain the
beam sizes. Those measurements were obtained with 3\,ms integration
time (in order to achieve similar SNR with less overall light
throughput), and so we compare them against the analysis of the
five-aperture interferometry in Table~\ref{table:comparison}. As for
the 5 apertures, the two-aperture SRI measurements employed 30 consecutive frames in each of four rotated positions of the mask. These rotated positions are required to perform the full 2D characterization in this method.

Table~\ref{table:comparison} also shows the expected results for
horizontal and vertical beam size and the tilt angle at the
synchrotron radiation extraction point obtained using Linear Optics
from Closed Orbits (LOCO, \cite{SAFRANEK199727}).  LOCO is a tool to
measure and correct the orbit of a circular accelerator. It takes into
account several measurements of the machine, such as current in magnets and BPM readings, and models them to estimate the expected beam behavior. The obtained result is not necessarily exact but indicative.
{We also list the beam size inferred from the emittance values measured by the two X-ray pinholes at ALBA~\cite{UI:IBIC22}, and the expected Twiss parameters at the source size of the NRA experiment.  }

Table~\ref{table:comparison} shows consistency within the errors between the 5-hole, 2-hole and pinhole beam size measurements. The SRI measurements are also within 4\% to 20\% of the LOCO modeling, depending on parameter. 

\begin{table}
  \centering 
\begin{tabular}{lcccc}
\toprule
Method & Major\,(\SI{}{\micro\metre}) & Minor\,(\SI{}{\micro\metre}) & Tilt Angle  \\
  \midrule
  2ap (3\,ms)    &  $61.7 \pm 1.5$ & $25.5 \pm 1.5$ & $16.6\degree$\\
  5ap (1\,ms)   &  $59.6 \pm 0.1$ & $23.8 \pm 0.5$ & $15.9\degree \pm 0.2\degree$\\
  LOCO  & 57.5 & 20.6 & 14.9$\degree$\\
  Pinhole  & 58.5 & 24.6 & 14.9$\degree$\\
\bottomrule
\end{tabular}
\caption{Comparison of beam size measurements by various methods. 2ap: two-aperture
  rotated mask. 5ap: the present five-aperture method. LOCO: ideal value based on the analysis of the magnet lattice of ALBA \cite{SAFRANEK199727}.  Pinhole: beam size inferred from the emittance calculation at the ALBA x-ray pinholes, although at a different location in the ring~\cite{UI:IBIC22}.
  }
\label{table:comparison}
\end{table}

\section{Random and Systematic Errors}\label{sec:rand-syst-errors}

{Our error estimates on measured quantities in
  Table~\ref{table:comparison} are empirical, based on the variance in
  a sequence of 30 measurements. In this section we look further at
  random and potential systematic errors in our technique.}

{The fundamental limit to accuracy comes from the photon
  counting statistics. These follow the Poisson distribution where the standard error for large count rates is the
  square root of the number of photons. Based on the measured counts
  in the visibilities, we estimate this error to be small, a relative
  error of $\le 0.2\%$. For reference, we have performed an analysis
  of the fitting process and find that a relative random error of
  $\sim 1\%$ in the coherences per frame would lead to the empirically
  measured scatter of the fitted beam quantities from the time
  series. }

{There are also a number of potential systematic errors which we
  estimate here together with our correction and mitigation strategy.}

\begin{figure}
\centering 
\centerline{\includegraphics[scale=0.18]{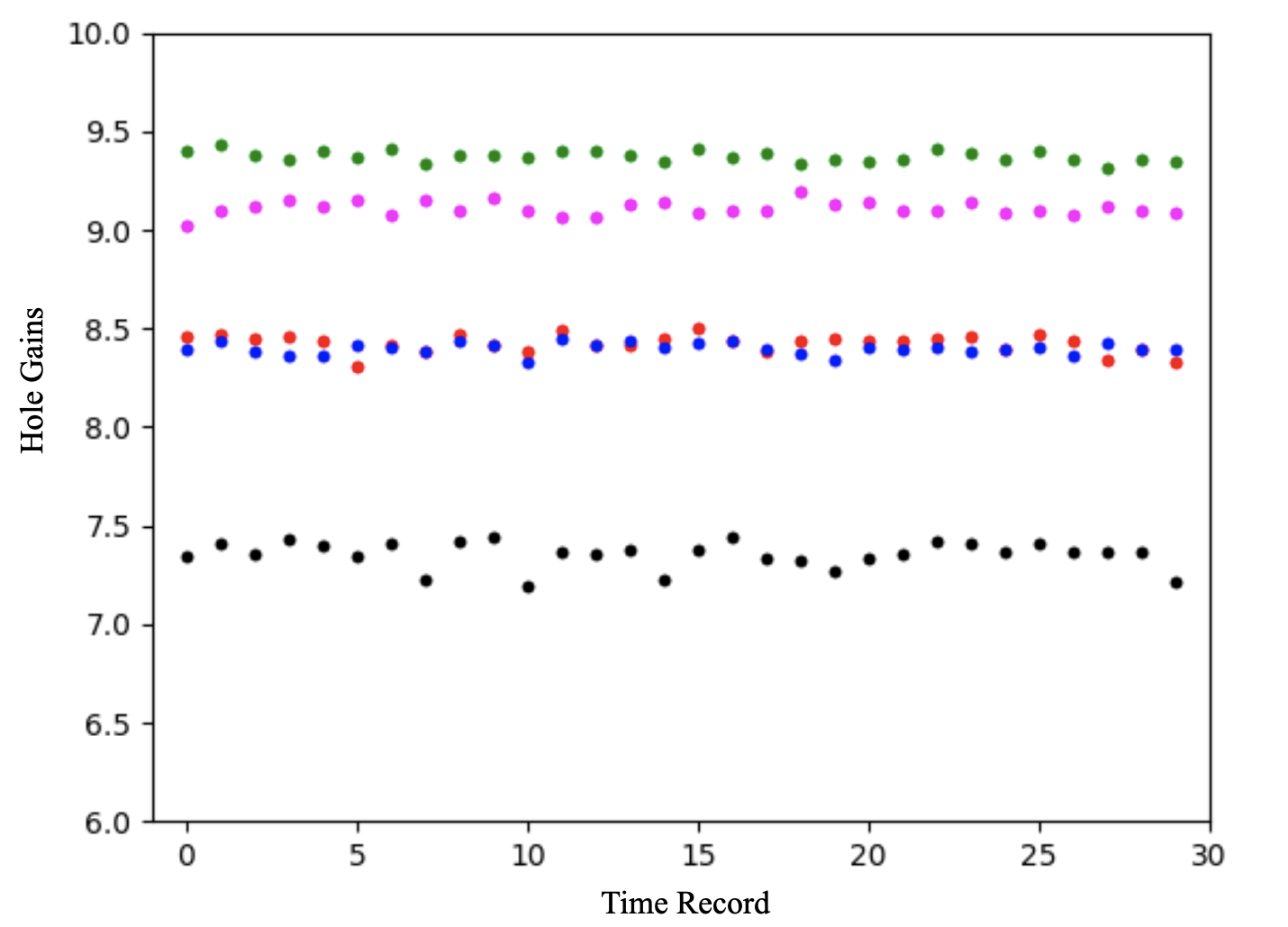}}
\caption{Solutions for the hole gain amplitudes for each hole for the 30 frames, as defined in equation 5. The units of the gains are in root(counts)/$10^3$, and derived such that the resulting coherences are normalized to unit peak. black = Ap0; red = Ap1; blue = Ap2; green = Ap3; Purple = Ap4. 
}
\label{fig:gain5H}
\end{figure}

{First, we demonstrate the importance of gain self-calibration
  to correct for the non-uniform illumination across the mask, as
  defined in Equation 5. The gains, $\|G_i\|$, derived for the 5 holes
  over the 30 time records are shown in Figure~\ref{fig:gain5H}.  The
  gains are stable to $\le 1\%$ over the 30 records, but the values
  vary from hole to hole by up to 25\% (which implies a $\sim 50\%$
  change in power of the illumination across the mask). This is due to
  unavoidable differences in the illumination from the beam across the
  mask, in part due to the inherent vertical intensity distribution of
  the synchrotron light and in part due to Fraunhoffer diffraction
  produced by various mirror edges.}

{We quantify the impact of this gain variation by an alternate
  fit for the source size without correcting the visibility amplitudes
  by the gains. In this analysis the derived source major axis is too
  large by typically 15\%, and the position angle and minor axis sizes
  are wrong by a factor 2 or more. The sensitivity of the minor axis
  to the gain corrections reflects the fact that this axis is only
  marginally resolved by our longest baseline, and hence true
  coherences are high, 90\% or more, such that even small corrections to
  the measured visibilities lead to substantial changes in fitted beam
  minor axis. Further, the agreement between the gain-corrected beam
  size and the expected beam size from other methods lends confidence
  that the gain correction process is valid.}

{The gains can alternatively be determined by closing in turn
  all but one of the holes on the mask and measuring the flux received
  at the CCD. Such a sequential measurement of total power received is
  sensitive to effects such as variation in the read noise of the CCD
  and the intensity of the synchrotron radiation. Furthermore the
  measurement would not be accurate after any appreciable change in
  the position or orientation of the mask since the illumination
  pattern on the mask has diffraction features on scales comparable to
  the hole sizes. This approach is hence less convenient for routine
  use but is a good way to check the approach in which the gains are
  fitted directly with the source and we are in the process of
  performing these tests at ALBA.}

{We also investigated how specific parameter choices in our
  approach affect the measured visibilities, which givens an
  illustration of the residual systematic error may be. The following
  were investigated:}
\begin{itemize}
\item {Not centering the image on the Airy disk before Fourier
    transforming to the $uv$ plane leads to changes in visibilities in
    range $\sim 1\% \textrm{--} 4\%.$, due to a phase ramp across the region of $uv$ plane over which the visibilities are measured. }
\item {Changing the radius for summing the complex visibilities
    in the $uv$ plane from 3 to 9 pixels changes the coherences by $\le 2\%$. }
\item {Changing the hole size from 5\,mm to 3\,mm raises the
    coherences in range $\sim 5 \textrm{--} 10\%$}
\item {Changing the frame time from 3ms to 1ms raises the coherences by $\sim 1\% \textrm{--} 8\%.$}
\item {Not removing a mean frame bias lowers coherences $\sim 2\%$.}
\end{itemize}
{The large impact of the hole size and integration time suggests
de-correlation on longer length and time scales, which could be due to
air turbulence in the lab or vibration of some parts of the instrument
or optical chain.  The impact of other parameters show that a careful
analysis and calibration is needed if the theoretical noise limit is
to be achieved.}

\begin{figure}
\centering 
\centerline{\includegraphics[scale=0.3]{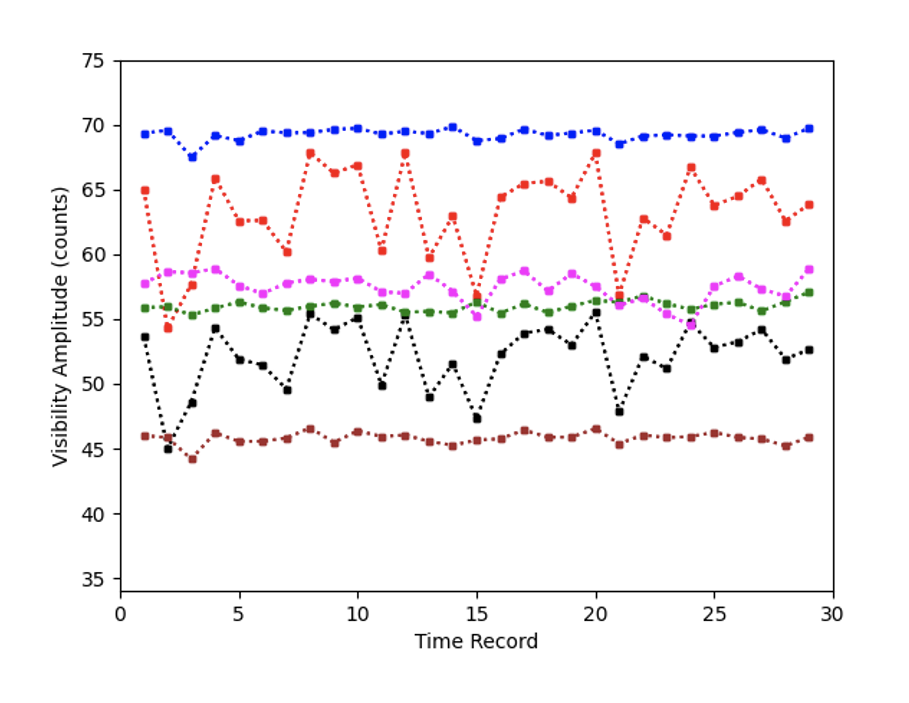}}
\caption{Time series of some of the visibility amplitudes for a 6 hole mask in which two of the visibilities (horizontal 16mm and vertical 16mm), are measured with redundant baselines (red and black curves), while the rest are for non-redundant baselines. The visibility amplitude axis is in units of $10^6$ counts.}
\label{fig:Redundancy}
\end{figure}

{Finally, we demonstrate the importance of using a non-redundant
  mask by repeating the experiment with redundant mask and comparing
  the results. For this test, a six hole mask was employed, identical
  to the current 5 hole mask, but including the fourth corner hole,
  such that the we now have two redundant baselines: the two 16mm
  horizontal baselines and two 16mm vertical
  baselines. Figure~\ref{fig:Redundancy} shows the results for six of
  the baselines, including the two redundant baselines. We find the
  RMS scatter on the redundant baselines is much higher, by a factor 3
  to 5, likely arising from time varying decoherence due to phase
  fluctuations, and that the mean amplitude is lower by about 5\% than
  expected using the non-redundant 5 hole mask. }

{
  The current experiment is simply a demonstration of the technique, which is apparently verified by the good agreement between the results from this technique with three other independent techniques for beam size measurements. We are continuing these investigations in order to improve the performance. For next-generation instruments, the ultimate resolution limit, $\delta\theta$, is set by the signal-to-noise ratio and the interferometric fringe spacing of the maximum baseline, $\rm B_{max}$, of the mask: $\rm \delta\theta \sim \theta_{Bmax}/(S/N)$, where $\rm \theta_{Bmax} \sim \lambda/B_{max}$ (see Fomalont in \cite{SIRAIIFomalont1999} equ 14-5). In our practical case at ALBA, for a typical $B_{\rm max}$=20~mm, $S/N$=0.2\%  and averaging 30 frames, the ultimate resolution becomes $\delta_X \sim$\SI{0.1}{\micro m}, consistent with best error estimate from variance between frames shown in Table~\ref{table:comparison}.
  }
  
{  We are in the process of improving our SRI technique through masks with more apertures over a larger area of the beam footprint on the mask, shorter integration times and hole size, and decreasing the measurement wavelength. Based on the various errors and uncertainties discussed above, we feel coherence measurement below 1\% of the true values are plausible. In general, future SRI optical systems would benefit from allowing for longer baselines (larger photon beam footprints on the mask), and shorter wavelengths, implying higher resolution. Extension of the technique to much shorter wavelengths, even to the EUV, is feasible, although focusing optics become more of a challenge, as does mask production. 
  }

\section{Conclusions}

The {particle} beam can be well characterised using a single interferogram using the emitted synchrotron radiation and a simple apparatus with no moving parts. 
The relatively large CCD image data set can be processed without iterative model-fitting to derive a much smaller data set of visibilities (in our case 10), which are then input into a fitting procedure. Importantly, this model fitting needs to fit for both the differential illumination of the apertures as well as for the coherences. The overall processing complexity is low and suitable for near-real time applications. 
The beam parameters and uncertainty in a single acquisition with a 5-hole non-redundant mask are comparable to those obtained with the two-aperture method employing multiple acquisitions with rotation of the aperture. The larger amount of light let through by the five apertures contributes to the efficiency of the technique.

{A key aspect of the technique is the use of non-redundant
  aperture mask which means that the visibility between each pair of
  apertures is measured separately. This means that distortions in the
  incoming wavefront at the aperture plane can be measured as function
  of position in the plane as well as a function of time (the latter
  by taking frames with short enough integration time). The spatial
  fluctuation of the incoming wavefront can be both in amplitude
  (which is the seen as the uneven illumination pattern on the mask)
  or in phase (e.g., due to atmospheric turbulence or optics vibrations). As discussed in
  Section~\ref{sec:rand-syst-errors} , both of these errors are
  significant. Although characterising a Gaussian function does not
  make use of the measured phases, it is still susceptible to phase
  errors across the aperture mask if visibilities with different phase
  errors are combined as they are in a redundant mask or filled
  aperture system.}

{Therefore by use of a non-redundant mask, and by making the
  individual apertures small enough and the integration times short
  enough, it is possible to  at least reduce the effects of both
  amplitude and phase errors which vary in position, time or a   combination of both.}
 In order for the model and the $\|G_i\|$ values to be well constrained, there must be more measured visibilities than the sum of the number of apertures and the number of model free parameters \cite{1980SPIE..231...18S}. This is satisfied with 5 apertures for the Gaussian model of the synchrotron beam. More complex models are possible, e.g., in astronomical applications it is also routine to iteratively build up a model when there is a significant number of measurements and the source plane can be assumed to sparse \cite{1980SPIE..231...18S}. In these cases, a larger number of apertures would be needed.

The technique presented can likewise be used for efficient and rapid
characterisation of other high-intensity light sources even after
propagation through a medium inducing phase and amplitude
fluctuations. Moreover, the complex visibilities and $G_i$ values
encode precise information on the optical performance of the
system. {For example, the static phases of $G_i$ are a
  direct measure of wavefront phase error accumulated in the optical
  system up to the aperture mask and hence can be used to infer the
  error in larger optical components. Slowly varying phases
  of $G_i$ after a change in experimental condition can be used to,
  for example, measure the deformation of the optical system due to
  thermal expansion/contraction. }

The technique presented here is a good demonstration of interdisciplinary
science, applying astronomical interferometric techniques to
accelerator physics.

\vskip 0.2in

{\bf Acknowledgments.} The National Radio Astronomy Observatory is a
facility of the National Science Foundation operated under cooperative
agreement by Associated Universities, Inc. Patent applied for: UK
Patent Application Number 2406928.8, USA Applications No. 63/648,303
(RL 8127.306.USPR) and No. 63/648,284 (RL 8127.035.GBPR).  {We
  thank two anonymous referees for their comments which have improved
  the paper presentation.}

\bibliography{refs}

\begin{thebibliography}{21}%
\makeatletter
\providecommand \@ifxundefined [1]{%
 \@ifx{#1\undefined}
}%
\providecommand \@ifnum [1]{%
 \ifnum #1\expandafter \@firstoftwo
 \else \expandafter \@secondoftwo
 \fi
}%
\providecommand \@ifx [1]{%
 \ifx #1\expandafter \@firstoftwo
 \else \expandafter \@secondoftwo
 \fi
}%
\providecommand \natexlab [1]{#1}%
\providecommand \enquote  [1]{``#1''}%
\providecommand \bibnamefont  [1]{#1}%
\providecommand \bibfnamefont [1]{#1}%
\providecommand \citenamefont [1]{#1}%
\providecommand \href@noop [0]{\@secondoftwo}%
\providecommand \href [0]{\begingroup \@sanitize@url \@href}%
\providecommand \@href[1]{\@@startlink{#1}\@@href}%
\providecommand \@@href[1]{\endgroup#1\@@endlink}%
\providecommand \@sanitize@url [0]{\catcode `\\12\catcode `\$12\catcode
  `\&12\catcode `\#12\catcode `\^12\catcode `\_12\catcode `\%12\relax}%
\providecommand \@@startlink[1]{}%
\providecommand \@@endlink[0]{}%
\providecommand \url  [0]{\begingroup\@sanitize@url \@url }%
\providecommand \@url [1]{\endgroup\@href {#1}{\urlprefix }}%
\providecommand \urlprefix  [0]{URL }%
\providecommand \Eprint [0]{\href }%
\providecommand \doibase [0]{https://doi.org/}%
\providecommand \selectlanguage [0]{\@gobble}%
\providecommand \bibinfo  [0]{\@secondoftwo}%
\providecommand \bibfield  [0]{\@secondoftwo}%
\providecommand \translation [1]{[#1]}%
\providecommand \BibitemOpen [0]{}%
\providecommand \bibitemStop [0]{}%
\providecommand \bibitemNoStop [0]{.\EOS\space}%
\providecommand \EOS [0]{\spacefactor3000\relax}%
\providecommand \BibitemShut  [1]{\csname bibitem#1\endcsname}%
\let\auto@bib@innerbib\@empty
\bibitem [{\citenamefont {Sands}(1969)}]{Sands:1969lzn}%
  \BibitemOpen
  \bibfield  {author} {\bibinfo {author} {\bibfnamefont {M.}~\bibnamefont
  {Sands}},\ }\bibfield  {title} {\bibinfo {title} {{The Physics of Electron
  Storage Rings: An Introduction}},\ }\href@noop {} {\bibfield  {journal}
  {\bibinfo  {journal} {Conf. Proc. C}\ }\textbf {\bibinfo {volume}
  {6906161}},\ \bibinfo {pages} {257} (\bibinfo {year} {1969})}\BibitemShut
  {NoStop}%
\bibitem [{\citenamefont {Kube}(2007)}]{Kube:Dipac07}%
  \BibitemOpen
  \bibfield  {author} {\bibinfo {author} {\bibfnamefont {G.}~\bibnamefont
  {Kube}},\ }\bibfield  {title} {\bibinfo {title} {{Review of Synchrotron
  Radiation Based Diagnostics for Transverse Profile Measurements}},\ }in\
  \href {https://accelconf.web.cern.ch/d07/papers/moo1a03.pdf} {\emph {\bibinfo
  {booktitle} {{Proceedings of DIPAC07, Venice (Italy)}}}}\ (\bibinfo {year}
  {2007})\BibitemShut {NoStop}%
\bibitem [{\citenamefont {Elleaume}\ \emph {et~al.}(1995)\citenamefont
  {Elleaume}, \citenamefont {Fortgang}, \citenamefont {Penel},\ and\
  \citenamefont {Tarazona}}]{Elleaume:km0006}%
  \BibitemOpen
  \bibfield  {author} {\bibinfo {author} {\bibfnamefont {P.}~\bibnamefont
  {Elleaume}}, \bibinfo {author} {\bibfnamefont {C.}~\bibnamefont {Fortgang}},
  \bibinfo {author} {\bibfnamefont {C.}~\bibnamefont {Penel}},\ and\ \bibinfo
  {author} {\bibfnamefont {E.}~\bibnamefont {Tarazona}},\ }\bibfield  {title}
  {\bibinfo {title} {{Measuring Beam Sizes and Ultra-Small Electron Emittances
  Using an X-ray Pinhole Camera}},\ }\href
  {https://doi.org/10.1107/S0909049595008685} {\bibfield  {journal} {\bibinfo
  {journal} {Journal of Synchrotron Radiation}\ }\textbf {\bibinfo {volume}
  {2}},\ \bibinfo {pages} {209} (\bibinfo {year} {1995})}\BibitemShut {NoStop}%
\bibitem [{\citenamefont {Trebushinin}\ \emph {et~al.}(2024)\citenamefont
  {Trebushinin}, \citenamefont {Geloni}, \citenamefont {Serkez}, \citenamefont
  {Khubbutdinov},\ and\ \citenamefont {Saldin}}]{PhysRevAccelBeams.27.032802}%
  \BibitemOpen
  \bibfield  {author} {\bibinfo {author} {\bibfnamefont {A.}~\bibnamefont
  {Trebushinin}}, \bibinfo {author} {\bibfnamefont {G.}~\bibnamefont {Geloni}},
  \bibinfo {author} {\bibfnamefont {S.}~\bibnamefont {Serkez}}, \bibinfo
  {author} {\bibfnamefont {R.}~\bibnamefont {Khubbutdinov}},\ and\ \bibinfo
  {author} {\bibfnamefont {E.}~\bibnamefont {Saldin}},\ }\bibfield  {title}
  {\bibinfo {title} {Pinhole camera for electron beam size diagnostic at
  storage ring with an ultralow emittance},\ }\href
  {https://doi.org/10.1103/PhysRevAccelBeams.27.032802} {\bibfield  {journal}
  {\bibinfo  {journal} {Phys. Rev. Accel. Beams}\ }\textbf {\bibinfo {volume}
  {27}},\ \bibinfo {pages} {032802} (\bibinfo {year} {2024})}\BibitemShut
  {NoStop}%
\bibitem [{\citenamefont {Mitsuhashi}(1998)}]{Mitsuhashi:1998em}%
  \BibitemOpen
  \bibfield  {author} {\bibinfo {author} {\bibfnamefont {T.}~\bibnamefont
  {Mitsuhashi}},\ }\bibfield  {title} {\bibinfo {title} {{Beam profile and size
  measurement by SR interferometers}},\ }in\ \href@noop {} {\emph {\bibinfo
  {booktitle} {{Joint US-CERN-Japan-Russia School on Particle Accelerators:
  Beam Measurement}}}}\ (\bibinfo {year} {1998})\ pp.\ \bibinfo {pages}
  {399--427}\BibitemShut {NoStop}%
\bibitem [{\citenamefont {Torino}\ and\ \citenamefont
  {Iriso}(2016)}]{PhysRevAccelBeams.19.122801}%
  \BibitemOpen
  \bibfield  {author} {\bibinfo {author} {\bibfnamefont {L.}~\bibnamefont
  {Torino}}\ and\ \bibinfo {author} {\bibfnamefont {U.}~\bibnamefont {Iriso}},\
  }\bibfield  {title} {\bibinfo {title} {Transverse beam profile reconstruction
  using synchrotron radiation interferometry},\ }\href
  {https://doi.org/10.1103/PhysRevAccelBeams.19.122801} {\bibfield  {journal}
  {\bibinfo  {journal} {Phys. Rev. Accel. Beams}\ }\textbf {\bibinfo {volume}
  {19}},\ \bibinfo {pages} {122801} (\bibinfo {year} {2016})}\BibitemShut
  {NoStop}%
\bibitem [{\citenamefont {Masaki}\ and\ \citenamefont
  {Takano}(2003)}]{Masaki:wr2006}%
  \BibitemOpen
  \bibfield  {author} {\bibinfo {author} {\bibfnamefont {M.}~\bibnamefont
  {Masaki}}\ and\ \bibinfo {author} {\bibfnamefont {S.}~\bibnamefont
  {Takano}},\ }\bibfield  {title} {\bibinfo {title} {{Two-dimensional visible
  synchrotron light interferometry for transverse beam-profile measurement at
  the SPring-8 storage ring}},\ }\href
  {https://doi.org/10.1107/S0909049503007106} {\bibfield  {journal} {\bibinfo
  {journal} {Journal of Synchrotron Radiation}\ }\textbf {\bibinfo {volume}
  {10}},\ \bibinfo {pages} {295} (\bibinfo {year} {2003})}\BibitemShut
  {NoStop}%
\bibitem [{\citenamefont {Carilli}\ \emph {et~al.}(2024)\citenamefont
  {Carilli}, \citenamefont {Nikolic}, \citenamefont {Torino}, \citenamefont
  {Iriso},\ and\ \citenamefont {Thyagarajan}}]{Carilli24}%
  \BibitemOpen
  \bibfield  {author} {\bibinfo {author} {\bibfnamefont {C.}~\bibnamefont
  {Carilli}}, \bibinfo {author} {\bibfnamefont {B.}~\bibnamefont {Nikolic}},
  \bibinfo {author} {\bibfnamefont {L.}~\bibnamefont {Torino}}, \bibinfo
  {author} {\bibfnamefont {U.}~\bibnamefont {Iriso}},\ and\ \bibinfo {author}
  {\bibfnamefont {N.}~\bibnamefont {Thyagarajan}},\ }\href@noop {} {\emph
  {\bibinfo {title} {{Deriving the size and shape of the ALBA synchrotron light
  source with optical aperture masking: technical choices}}}},\ \bibinfo {type}
  {Tech. Rep.}\ (\bibinfo  {institution} {ALBA},\ \bibinfo {year}
  {2024})\BibitemShut {NoStop}%
\bibitem [{\citenamefont {Novokshonov}\ \emph {et~al.}(2017)\citenamefont
  {Novokshonov}, \citenamefont {Kube},\ and\ \citenamefont
  {Potylitsyn}}]{Novokshonov:2017qsa}%
  \BibitemOpen
  \bibfield  {author} {\bibinfo {author} {\bibfnamefont {A.}~\bibnamefont
  {Novokshonov}}, \bibinfo {author} {\bibfnamefont {G.}~\bibnamefont {Kube}},\
  and\ \bibinfo {author} {\bibfnamefont {A.}~\bibnamefont {Potylitsyn}},\
  }\bibfield  {title} {\bibinfo {title} {{Two-Dimensional Synchrotron Radiation
  Interferometry at PETRA III}},\ }in\ \href
  {https://doi.org/10.18429/JACoW-IPAC2017-MOPAB042} {\emph {\bibinfo
  {booktitle} {{8th International Particle Accelerator Conference}}}}\
  (\bibinfo {year} {2017})\BibitemShut {NoStop}%
\bibitem [{\citenamefont {Li}\ \emph {et~al.}(2023)\citenamefont {Li},
  \citenamefont {Lu}, \citenamefont {Lu},\ and\ \citenamefont {Wang}}]{Li:23}%
  \BibitemOpen
  \bibfield  {author} {\bibinfo {author} {\bibfnamefont {Q.}~\bibnamefont
  {Li}}, \bibinfo {author} {\bibfnamefont {Y.}~\bibnamefont {Lu}}, \bibinfo
  {author} {\bibfnamefont {Y.}~\bibnamefont {Lu}},\ and\ \bibinfo {author}
  {\bibfnamefont {P.}~\bibnamefont {Wang}},\ }\bibfield  {title} {\bibinfo
  {title} {Two-dimensional spatial coherence measurement of x-ray sources using
  aperture array mask},\ }\href {https://doi.org/10.1364/OE.503171} {\bibfield
  {journal} {\bibinfo  {journal} {Opt. Express}\ }\textbf {\bibinfo {volume}
  {31}},\ \bibinfo {pages} {36304} (\bibinfo {year} {2023})}\BibitemShut
  {NoStop}%
\bibitem [{\citenamefont {Siano}\ \emph {et~al.}(2022)\citenamefont {Siano},
  \citenamefont {Paroli}, \citenamefont {Potenza}, \citenamefont {Teruzzi},
  \citenamefont {Iriso}, \citenamefont {Nosych}, \citenamefont {Solano},
  \citenamefont {Torino}, \citenamefont {Butti}, \citenamefont {Goetz},
  \citenamefont {Lefevre}, \citenamefont {Mazzoni},\ and\ \citenamefont
  {Trad}}]{PhysRevAccelBeams.25.052801}%
  \BibitemOpen
  \bibfield  {author} {\bibinfo {author} {\bibfnamefont {M.}~\bibnamefont
  {Siano}}, \bibinfo {author} {\bibfnamefont {B.}~\bibnamefont {Paroli}},
  \bibinfo {author} {\bibfnamefont {M.~A.~C.}\ \bibnamefont {Potenza}},
  \bibinfo {author} {\bibfnamefont {L.}~\bibnamefont {Teruzzi}}, \bibinfo
  {author} {\bibfnamefont {U.}~\bibnamefont {Iriso}}, \bibinfo {author}
  {\bibfnamefont {A.~A.}\ \bibnamefont {Nosych}}, \bibinfo {author}
  {\bibfnamefont {E.}~\bibnamefont {Solano}}, \bibinfo {author} {\bibfnamefont
  {L.}~\bibnamefont {Torino}}, \bibinfo {author} {\bibfnamefont
  {D.}~\bibnamefont {Butti}}, \bibinfo {author} {\bibfnamefont
  {A.}~\bibnamefont {Goetz}}, \bibinfo {author} {\bibfnamefont
  {T.}~\bibnamefont {Lefevre}}, \bibinfo {author} {\bibfnamefont
  {S.}~\bibnamefont {Mazzoni}},\ and\ \bibinfo {author} {\bibfnamefont
  {G.}~\bibnamefont {Trad}},\ }\bibfield  {title} {\bibinfo {title}
  {Two-dimensional electron beam size measurements with x-ray heterodyne near
  field speckles},\ }\href
  {https://doi.org/10.1103/PhysRevAccelBeams.25.052801} {\bibfield  {journal}
  {\bibinfo  {journal} {Phys. Rev. Accel. Beams}\ }\textbf {\bibinfo {volume}
  {25}},\ \bibinfo {pages} {052801} (\bibinfo {year} {2022})}\BibitemShut
  {NoStop}%
\bibitem [{\citenamefont {{Labeyrie}}(1970)}]{1970A&A.....6...85L}%
  \BibitemOpen
  \bibfield  {author} {\bibinfo {author} {\bibfnamefont {A.}~\bibnamefont
  {{Labeyrie}}},\ }\bibfield  {title} {\bibinfo {title} {{Attainment of
  Diffraction Limited Resolution in Large Telescopes by Fourier Analysing
  Speckle Patterns in Star Images}},\ }\href@noop {} {\bibfield  {journal}
  {\bibinfo  {journal} {\aap}\ }\textbf {\bibinfo {volume} {6}},\ \bibinfo
  {pages} {85} (\bibinfo {year} {1970})}\BibitemShut {NoStop}%
\bibitem [{\citenamefont {Labat}\ \emph {et~al.}(2023)\citenamefont {Labat},
  \citenamefont {Chubar}, \citenamefont {Breunlin}, \citenamefont {Hubert},\
  and\ \citenamefont {Andersson}}]{PhysRevLett.131.185001}%
  \BibitemOpen
  \bibfield  {author} {\bibinfo {author} {\bibfnamefont {M.}~\bibnamefont
  {Labat}}, \bibinfo {author} {\bibfnamefont {O.}~\bibnamefont {Chubar}},
  \bibinfo {author} {\bibfnamefont {J.}~\bibnamefont {Breunlin}}, \bibinfo
  {author} {\bibfnamefont {N.}~\bibnamefont {Hubert}},\ and\ \bibinfo {author}
  {\bibfnamefont {A.}~\bibnamefont {Andersson}},\ }\bibfield  {title} {\bibinfo
  {title} {Bending magnet synchrotron radiation imaging with large orbital
  collection angles},\ }\href {https://doi.org/10.1103/PhysRevLett.131.185001}
  {\bibfield  {journal} {\bibinfo  {journal} {Phys. Rev. Lett.}\ }\textbf
  {\bibinfo {volume} {131}},\ \bibinfo {pages} {185001} (\bibinfo {year}
  {2023})}\BibitemShut {NoStop}%
\bibitem [{\citenamefont {Curcio}\ \emph {et~al.}(2024)\citenamefont {Curcio}
  \emph {et~al.}}]{Curcio2024}%
  \BibitemOpen
  \bibfield  {author} {\bibinfo {author} {\bibnamefont {Curcio}} \emph
  {et~al.},\ }\bibfield  {title} {\bibinfo {title} {Reconstruction of lateral
  coherence and 2d emittance in plasma betatron x-ray sources},\ }\href@noop {}
  {\bibfield  {journal} {\bibinfo  {journal} {Scientific Reports}\ }\textbf
  {\bibinfo {volume} {14}} (\bibinfo {year} {2024})}\BibitemShut {NoStop}%
\bibitem [{\citenamefont {{Haniff}}\ \emph {et~al.}(1987)\citenamefont
  {{Haniff}}, \citenamefont {{Mackay}}, \citenamefont {{Titterington}},
  \citenamefont {{Sivia}},\ and\ \citenamefont
  {{Baldwin}}}]{1987Natur.328..694H}%
  \BibitemOpen
  \bibfield  {author} {\bibinfo {author} {\bibfnamefont {C.~A.}\ \bibnamefont
  {{Haniff}}}, \bibinfo {author} {\bibfnamefont {C.~D.}\ \bibnamefont
  {{Mackay}}}, \bibinfo {author} {\bibfnamefont {D.~J.}\ \bibnamefont
  {{Titterington}}}, \bibinfo {author} {\bibfnamefont {D.}~\bibnamefont
  {{Sivia}}},\ and\ \bibinfo {author} {\bibfnamefont {J.~E.}\ \bibnamefont
  {{Baldwin}}},\ }\bibfield  {title} {\bibinfo {title} {{The first images from
  optical aperture synthesis}},\ }\href {https://doi.org/10.1038/328694a0}
  {\bibfield  {journal} {\bibinfo  {journal} {\nat}\ }\textbf {\bibinfo
  {volume} {328}},\ \bibinfo {pages} {694} (\bibinfo {year}
  {1987})}\BibitemShut {NoStop}%
\bibitem [{\citenamefont {{Schwab}}(1980)}]{1980SPIE..231...18S}%
  \BibitemOpen
  \bibfield  {author} {\bibinfo {author} {\bibfnamefont {F.~R.}\ \bibnamefont
  {{Schwab}}},\ }\bibfield  {title} {\bibinfo {title} {{Processing of
  three-dimensional data}},\ }in\ \href {https://doi.org/10.1117/12.958828}
  {\emph {\bibinfo {booktitle} {1980 International Optical Computing Conference
  I}}},\ \bibinfo {series} {Society of Photo-Optical Instrumentation Engineers
  (SPIE) Conference Series}, Vol.\ \bibinfo {volume} {231},\ \bibinfo {editor}
  {edited by\ \bibinfo {editor} {\bibfnamefont {W.~T.}\ \bibnamefont
  {{Rhodes}}}}\ (\bibinfo {year} {1980})\ p.~\bibinfo {pages} {18}\BibitemShut
  {NoStop}%
\bibitem [{\citenamefont {Iriso}\ and\ \citenamefont
  {Fernandez}(2011)}]{AAD-FE-DI-VMIR-01}%
  \BibitemOpen
  \bibfield  {author} {\bibinfo {author} {\bibfnamefont {U.}~\bibnamefont
  {Iriso}}\ and\ \bibinfo {author} {\bibfnamefont {F.}~\bibnamefont
  {Fernandez}},\ }in\ \href@noop {} {\emph {\bibinfo {booktitle} {{ ALBA
  Project Document Report No. AAD-FE-DI-VMIR-01}}}}\ (\bibinfo {year}
  {2011})\BibitemShut {NoStop}%
\bibitem [{\citenamefont {Gonz\'{a}lez}\ and\ \citenamefont
  {Mej\'{i}a}(2011)}]{Gonzalez:11}%
  \BibitemOpen
  \bibfield  {author} {\bibinfo {author} {\bibfnamefont {A.~I.}\ \bibnamefont
  {Gonz\'{a}lez}}\ and\ \bibinfo {author} {\bibfnamefont {Y.}~\bibnamefont
  {Mej\'{i}a}},\ }\bibfield  {title} {\bibinfo {title} {Nonredundant array of
  apertures to measure the spatial coherence in two dimensions with only one
  interferogram},\ }\href {https://doi.org/10.1364/JOSAA.28.001107} {\bibfield
  {journal} {\bibinfo  {journal} {J. Opt. Soc. Am. A}\ }\textbf {\bibinfo
  {volume} {28}},\ \bibinfo {pages} {1107} (\bibinfo {year}
  {2011})}\BibitemShut {NoStop}%
\bibitem [{\citenamefont {Safranek}(1997)}]{SAFRANEK199727}%
  \BibitemOpen
  \bibfield  {author} {\bibinfo {author} {\bibfnamefont {J.}~\bibnamefont
  {Safranek}},\ }\bibfield  {title} {\bibinfo {title} {Experimental
  determination of storage ring optics using orbit response measurements},\
  }\href {https://doi.org/https://doi.org/10.1016/S0168-9002(97)00309-4}
  {\bibfield  {journal} {\bibinfo  {journal} {Nuclear Instruments and Methods
  in Physics Research Section A: Accelerators, Spectrometers, Detectors and
  Associated Equipment}\ }\textbf {\bibinfo {volume} {388}},\ \bibinfo {pages}
  {27} (\bibinfo {year} {1997})}\BibitemShut {NoStop}%
\bibitem [{\citenamefont {Iriso}\ \emph {et~al.}(2022)\citenamefont {Iriso},
  \citenamefont {Marti}, \citenamefont {Nosych}, \citenamefont {Cazorla},\ and\
  \citenamefont {Mases}}]{UI:IBIC22}%
  \BibitemOpen
  \bibfield  {author} {\bibinfo {author} {\bibfnamefont {U.}~\bibnamefont
  {Iriso}}, \bibinfo {author} {\bibfnamefont {Z.}~\bibnamefont {Marti}},
  \bibinfo {author} {\bibfnamefont {A.}~\bibnamefont {Nosych}}, \bibinfo
  {author} {\bibfnamefont {A.}~\bibnamefont {Cazorla}},\ and\ \bibinfo {author}
  {\bibfnamefont {I.}~\bibnamefont {Mases}},\ }\bibfield  {title} {\bibinfo
  {title} {{PSF Characterization of the ALBA X-Ray Pinholes}},\ }in\ \href
  {https://accelconf.web.cern.ch/ibic2022/papers/wep16.pdf} {\emph {\bibinfo
  {booktitle} {Proceeding of IBIC2022}}}\ (\bibinfo {year} {2022})\BibitemShut
  {NoStop}%
\bibitem [{\citenamefont {Taylor}\ and\ \citenamefont
  {Perley}(1999)}]{SIRAIIFomalont1999}%
  \BibitemOpen
  \bibinfo {editor} {\bibfnamefont {C.~L.}\ \bibnamefont {Taylor},
  \bibfnamefont {G.~B.and~Carilli}}\ and\ \bibinfo {editor} {\bibfnamefont
  {R.~A.}\ \bibnamefont {Perley}},\ eds.,\ \href@noop {} {\emph {\bibinfo
  {title} {Synthesis Imaging in Radio Astronomy II}}},\ Vol.\ \bibinfo {volume}
  {180}\ (\bibinfo  {publisher} {ASP Conference Series},\ \bibinfo {year}
  {1999})\BibitemShut {NoStop}%
\end{thebibliography}%
\end{document}